# Improving fitness:

# Mapping research priorities against societal needs on obesity

Lorenzo Cassi[1], Agénor Lahatte[2], Ismael Rafols[3], Pierre Sautier[4] and Élisabeth de Turckheim[5]

[1] *lorenzo.cassi@uni-paris1.fr*
Observatoire des Sciences et Techniques (HCERES-OST)
Paris School of Economics, University Paris 1, Paris, France

[2] *agenor.lahatte@hceres.fr*
Observatoire des Sciences et Techniques (HCERES-OST), Paris, France

[3] *i.rafols@ingenio.upv.es*
Ingenio (CSIC-UPV), Universitat Politècnica de València, València, Spain,
CWTS, University of Leiden, Leiden, The Netherlands
SPRU (Science and Technology Policy Research), University of Sussex, Brighton, UK

[4] *pr.sautier@gmail.com*
Observatoire des Sciences et Techniques (HCERES-OST), Paris, France
Ingenio (CSIC-UPV), Universitat Politècnica de València, València, Spain

[5] *elisabeth.deturckheim@almacha.org*
*Observatoire des Sciences et Techniques (HCERES-OST), Paris, France*
*INRA, Délégation à l'évaluation, Paris, France*

## Abstract

Science policy is increasingly shifting towards an emphasis in societal problems or grand challenges. As a result, new evaluative tools are needed to help assess not only the knowledge production side of research programmes or organisations, but also the articulation of research agendas with societal needs. In this paper, we present an exploratory investigation of science supply and societal needs on the grand challenge of obesity - an emerging health problem with enormous social costs. We illustrate a potential approach that uses topic modelling to explore: (a) how scientific publications can be used to describe existing priorities in science production; (b) how policy records (in this case here questions posed in the European parliament) can be used as an instance of mapping discourse of social needs; (c) how the comparison between the two may show (mis)alignments between societal concerns and scientific outputs. While this is a technical exercise, we propose that this type of mapping methods can be useful to domain experts for informing strategic planning and evaluation in funding agencies.

## 1. Introduction

Assessing the contribution of research to address complex global problems or grand challenges - such as climate change, food security, poverty reduction, the risk of global pandemics - has become increasingly important in science policy as governments are under pressure to justify and legitimise their spending in research (Swedish Presidency of the European Union, 2009).

Conventional bibliometric techniques have been successful in providing tools that allow estimating production and research performance of scientific fields (Moed, 2005). In fact, they have been so successful, that their use has been stretched to contexts or areas beyond their validity or they have resulted in problematic or perverse incentives (Weingart, 2005; Hicks et al., 2015). However, addressing a societal problem does not only (or necessarily) require improving the production and quality of research on that problem. Conducting a lot of research of the highest quality about part of the knowledge base (e.g. electricity generation) is not enough if other parts of the knowledge are not achieved (e.g. electricity distribution). Addressing societal problems requires to link and potentially to coordinate of a variety of stakeholders with different areas of expertise and pursuing diverse research avenues (Ely, Van Zwanenberg & Stirling, 2014). Assessing scientific production and quality is not enough. Tools that help assess or manage the types of research topics, the types of actors involved and their relationships are thus needed.

In this article we explore a mapping methods to help identify research topics relevant for a societal challenge, in this case obesity. We believe these methods should be used interactively in close collaboration with domain experts as part of a large methodology that includes deliberation with diverse expertise and stakeholders. Obesity is an interesting issue for this exercise because it is a serious condition in which different types of policy interventions can be prioritised (PorGrow project, Millstone et al. 2006).

Obesity is “a critical global issue”. It has been considered as a disease by WHO since 1948 and its the global burden of disease has been highlighted since 1997 (James, 2008). Nearly 30 per cent of world population is estimated to be overweight or obese (Dobbs, Sawers, Thompson et al. 2014) and the current estimations predict that “if these trends continue, by 2025, global obesity prevalence will reach 18% in men and surpass 21% in women; severe obesity will surpass 6% in men and 9% in women” (NCD-RiskC, 2016). Moreover, obesity is a global issue since it concerns both developed and developing countries.

Facing this epidemic requires a systemic set of interventions that address the different issues related with obesity: biological and human metabolism; social, economic and family environments; education and life style; food supply and agribusiness policies, etc. Therefore measures should involve various types of actors, such as policy makers, educational institutions and community associations, media, firms, or restaurants and food distributors. However, the understanding and the empirical evidence available from research about the different facets and their interactions is very challenging. Therefore important investments on research on these topics are necessary and, at the same time, implementation of effective policy interventions should not be delayed (Dobbs, Sawers, Thompson et al. 2014). Because of this complexity, the analysis of the alignment - or the lack thereof - between perceived social needs and ongoing research priorities about obesity is a relevant case study for the methodology proposed in this paper.

We base the approach on Sarewitz and Pielke's (2007) concept of alignment (or lack of thereof) between the science produced (the knowledge *supply)* – and what is required to satisfy social needs (the knowledge *demand)*. Since research is conducted in conditions of highly incomplete knowledge, uncertainty plays an important role in scientific advancement – which is why it is important to keep a diversity of available options in research portfolios (Stirling, 2007; Wallace and Rafols, 2015). However, it is well documented that certain research options are much better aligned to specific desired outcomes (Sarewitz, 1996, pp. 31–49). For example, veterinary epidemiology is much more likely to help address avian influenza than high energy physics; and in the foreseeable future, solar energy from photovoltaic cells is more likely to help African farmers than nuclear fusion technology. Thus, the "supply" side of research should not only consider the quantity or quality of research, but also about the type of outcomes expected from a given research line.

The 'best' choices on the side of desired outcomes (knowledge demand) should be plural as well, because a given problem can be addressed with a variety of technologies. Which of this technologies is 'best' is uncertain (since they are not yet ready), and ambiguous (because different social actors may differ on their preferences). For example, some corporations may prefer nuclear energy whereas some environmentalist movement prefers solar energy. Or, in obesity, some actors may prefer investing on improved drugs, other on healthcare systems, rather than fostering changes in life styles or food production. It is therefore important to consider these ambiguities and identify the plurality and contested nature of desired outcomes. Public deliberation exercises are a way for stakeholders to make explicit their outcome preferences, which are likely to differ on the basis of divergent underlying values (Ely, Van Zwanenberg & Stirling, 2014). Furthermore, it is necessary to take into account various ways of articulating specific research options for achieving specific outcomes.

The research options that are prioritised in practice at a given moment reflect a variety of pressures and interests (a distributed governance) that favours certain research options, due for instance to power gradients in academia among disciplinary groups, and certain types of solutions, due to differences in economic incentives, e.g. pharmacological over life style changes (Wallace and Rafols, 2016). The distribution of resources over research options reflects a specific political economy of science, in particular the power asymmetries in the stakeholders involved (e.g. between energy utilities and consumers or between pharmaceutical companies and patients) (Tyfield, 2012).

The picture is thus one of multiple choices or options on both the supply side (research landscape) and the demand side (solutions/outcomes landscape): there are diverse research options that are variously related with different desired outcomes. Different stakeholders have different preferences, some directly on research options (e.g. disciplinary associations); other on outcomes (firms or users).

Ultimately, we aim to develop a methodology to explore the alignment or misalignment between science supply and societal needs or demands that will consist of a science mapping exercises aimed at informing stakeholder workshops for priority setting. The goal of this paper is to experiment with a mapping methodology. We do so, first, mapping the scientific supply via the research landscape of obesity as defined by topic modelling of publications titles and abstracts and, second, mapping social demand according to political discourse in the European parliament.

The results of a semantic mapping exercise are highly contingent on three factors: (i) choice of data source or database; (ii) the method used for delineating the corpus of the problem; (iii) the algorithms that define the cognitive categories of the corpus– which from a research portfolio perspective are the research options. Let us stress that all these choices may have a large effect on the results. This means that the maps presented are specific and partial representations of obesity research, and that other, complementary and/or contrasting representations are possible and equally legitimate.

Here, we use Web of Science and PubMed as data source of scientific publications. Notice that science supply could be drawn from more comprehensive databases, particularly regarding developing countries, such as CABI (specialised in agriculture and global health), which might lead to serious differences, as shown by Rafols, Ciarli and Chavarro (2015). Also science supply could be represented as well by many other type of data grants' abstracts rather than publications which might lead to a different configuration, as Talley et al. (2011) did for the US National Institutes of Health.

Topic delineation is controversial for issues such as obesity. The question is: what should be considered as research that matters to obesity? Only the publications that directly relate to obesity? Or also the publications that touch upon the issue more indirectly such as diet, physical exercise or public transportation? The answer is possibly a grey scale. As method for delineation, we borrowed an idea from Milanes, Noyons and de Faria, (2016) that considers as publications relevant to obesity all those that belong to publication micro-clusters with a relatively high percentage of obesity research.

Topic modelling only relies on textual information and it has been presented as an efficient method to extract the thematic structure of sets of non academic documents as web pages or press articles (Di Maggio, Nag and Blei, 2013; Klavans and Boyack, 2014). It is therefore a relevant method to extract the issues raised in questions by Members of the European Parliament (MEPs) about obesity. We also explore topic modelling as a method to identify research options from the corpus of publications related to obesity. This method does not make use of citation information that is available for scientific publications and has been successfully used for mapping scientific domains and partitioning large sets of publications. Large scale studies showed that text based metrics – where similarity between documents is calculated with terms frequencies – allow to map academic corpora when compared with citation based metrics (Boyack and Klavans, 2010; Boyack, Newmann, Duhon et al. 2011). In this paper, we do not consider other mapping approaches and we neither compare them with topic modelling. We leave this issue for a further work and here, we exploit the fact that the two corpora (publications and parliamentary questions) can be described with the same type of objects (i.e. topics) to tentatively compare them, despite their different nature (science vs politics) and size.

The paper is organised as follows. The second section presents the data used to capture both the supply and the demand sides and introduces topic modelling. The next three sections present the main results: first the topic modelling of research activities, secondly the topic modelling of EP political discourse, and finally the comparison between the two in terms of alignment or misalignment of their thematic. In the sixth section, we zoom in for a more grained description of topics involving social and environmental aspects. A discussion section comments the methodology choice in this paper and provides final remarks and directions for future work.

## 2. Data and methods

*Science supply data*

In this subsection we present the data used to represent supply (i.e. research activities) and in the next subsection data for capturing on perspective on the diversity of the demand (i.e. societal needs).

In order to define the relevant corpus for the research activity on obesity, we follow a two step method. First, we retrieve all publications MeSH term matching the search *obes** in MEDLINE/PubMed during the 2000-2013 period. This search was performed on October 16, 2014 and it returned 87,315 records.

Then, we launched *medlineR*, a routine based on the R language that allows the user to match data from Medline/PubMed with records indexed in the ISI Web of Science (WoS) database (Rotolo and Leydesdorff, 2015). The routine identified 71,055 WoS records (WoS core collections), with 'article' or 'review' as document types.

We then use the same search directly on the WoS Core Collection which returns 135,349 documents and we define the original corpus as the union of these two sets. At this stage, the original corpus consists in 147,322 documents where 59,079 of them were retrieved by both queries.

Second, we use a classification system generated by Waltman and van Eck at CWTS to identify clusters of publications related to obesity. This classification system is obtained from a publication-level clustering algorithm based on direct citations (Waltman and van Eck, 2012). Obesity publications appear at least once in 4,718 micro-clusters out of 32,466 micro-clusters of the whole classification. We enriched the original corpus with the whole clusters having at least a percentage $\alpha$ of publications tagged as obesity by the WoS query. Three levels of $\alpha$ were selected (50%, 30% and 10%) leading to three corpora: A, B and C. Finally, in order to use topic modelling on these corpora (which relies on abstracts), documents without an abstract were removed from these sets. This step reduces the original corpus of 147,322 documents to a corpus denoted O of 133,731 documents.

This process provided three extended corpora inflating the original corpus O. These corpora not only cover research aimed at tackling obesity but also, through references and citations, other knowledge related to obesity as for instance basic research in biology, medical research on other diseases of which obesity is a risk factor. As a benchmark of research that is strongly focused on obesity we consider a sub-corpus - denoted T - of publications where the term "obes*" is present in the title. This selects a very small subset of O with only 36,436 documents.

Table 1: Size of the five science corpora

| CORPUS NAME | MINIMAL % OF OBESITY PAPERS IN CLUSTERS | NUMBER OF CLUSTERS INCLUDED | CORPUS SIZE |
|---|---|---|---|
| T | obes* in title | 0 | 36,436 |
| O | obes* in title+abstract | 0 | 133,731 |
| A | 50% | 99 | 148,879 |
| B | 30% | 207 | 176,322 |
| C | 10% | 553 | 287,428 |

Obesity research increased a lot over time: in our largest corpus C, the average number of publications per year is 12,000 publications for the period 2002-2004, 21,000 in 2005-2009 and 30,000 in 2010-2013.

The disciplinary distribution of the corpora shows few differences (Fig.1). We use the Observatoire des Sciences et Techniques (OST) classification of journals into 33 disciplines based on WoS categories[1]. Nutrition & Endocrinology discipline has the largest share in any of the five corpora. It is more represented in corpus T than in corpus O and when clusters with a small coverage level are aggregated (corpus C). This is also true for Surgery & Gastroenterology & Urology. These disciplines are likely to provide a strict medical definition of obesity. On the contrary, the proportion of papers in Biochemistry and in Neuroscience increases when more clusters are added to the corpus.

[1] www.obs-ost.fr/sites/default/files/nomenclatures_disciplinaires_0.pdf

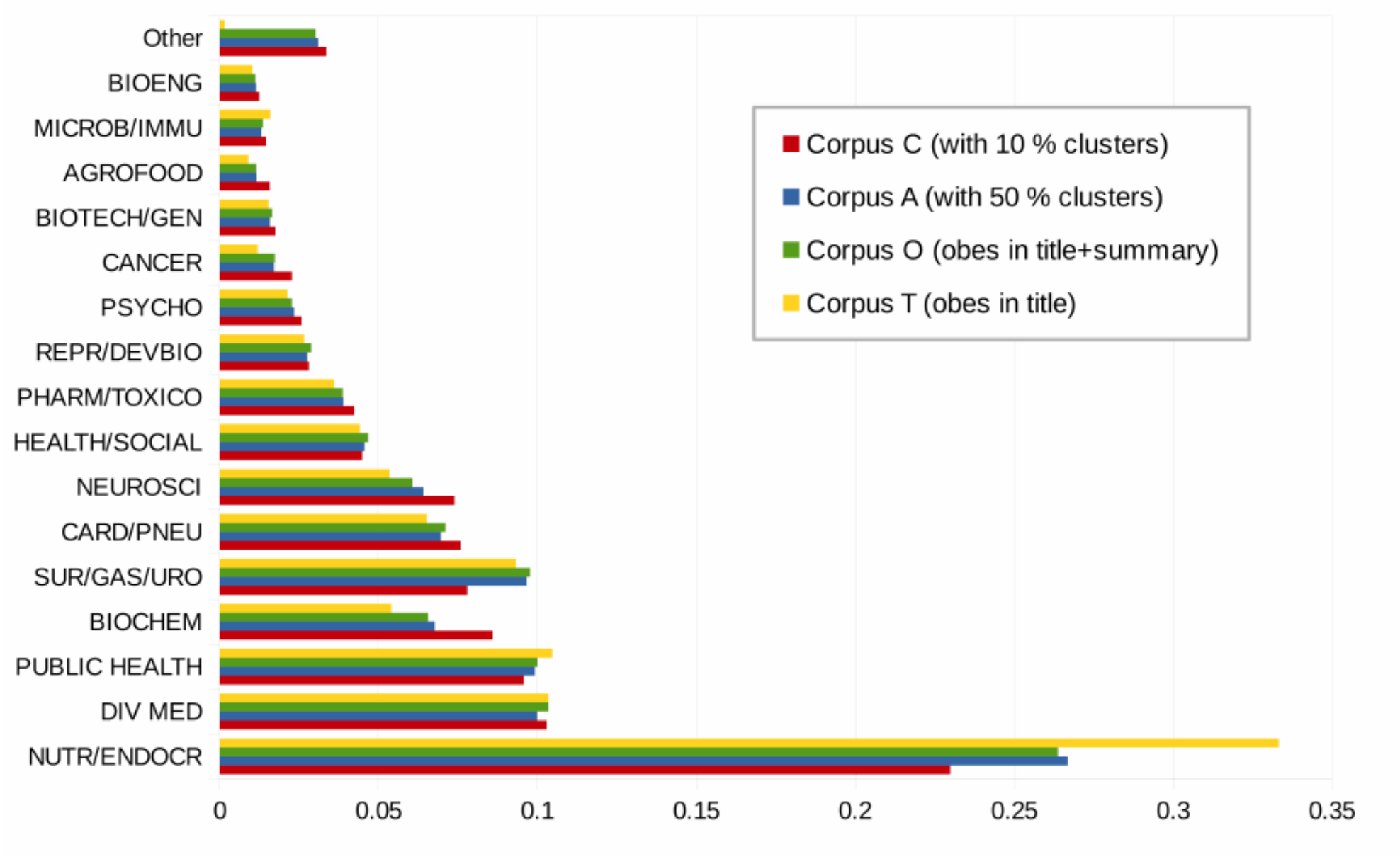

**Fig. 1. Discipline proportions of the 16 most represented OST disciplines in corpus C**. Non represented disciplines represent less than 0.4% of corpus C. The disciplinary profile of corpus B (with 30% clusters, not shown) is very much similar to that of corpus A.

For the further analysis, we choose to focus only on the largest corpus (i.e. corpus C). This choice responds to two reasons. On the one hand, using the largest corpus allows us to be inclusive and consequently to deal with a wide variety of research perspectives. This ensures a high recall which is necessary as the aim is to identify a possible lack of diversity in research approaches. On the other hand, it seems that being very inclusive does not add to much noise because the distribution of disciplines is not deeply modified in this process.

*Societal demands data*

Capturing the societal demands or needs requires exploring a wide range of sources to obtain an overview of the diversity of perspectives and expectations that social groups can support. Among others, press articles, web pages, or posts on social media are useful sources. For instance, these sources have been exploited by the Complex System Institute Paris Île de France (ISCPIF) to build a *Climate Tweetoscope* which analyses news on the Web with Twitter as a proxy to provide a map of questions and debates on climatic change (Chavalarias, Castillo and Panahi, 2015).

In principle, the description of the demand side should take into account the variety of actors, perspectives and desired outcomes. In this study, as a first technical test, we choose to analyse questions from Members of European Parliament (MEPs) to the European Commission. We understand that parliamentary question can be a proxy for societal demands because, as elected representatives, MEPs are expected to report the issues and needs of various stakeholders, lobbyists and social groups. According to this, their questions can be interpreted as a specific instance of social demands and needs. MEPs' questions are easily available, they are quite homogeneous texts in term of size (number of words) and vocabulary, what simplifies their treatment and analysis. We therefore retrieved all the questions (oral or written) asked during the Seventh Parliament term (2009-2014) and the beginning of the current term (until February 15, 2015) which contain the word obes* in the text[2]. This query harvested 222 questions. Questions are short texts (30 to 400 words), pointing to either the general issue of obesity or to particular causes or consequences of obesity. Facts reported are often based on a citation from a report or from a scientific publication and introduce focused queries about the actions planned or carried out by the European Commission to address a given issue.

*Method*

A topic model approach takes a collection of texts as input and shows hidden thematic structures as a set of topics (recurring themes that are discussed in the collection) and the degree to which each document exhibits those topics. Unlike clustering methods aiming at a partition of documents into clusters (or communities) of aiming at a partition of terms, topic modelling assumes that a document may address multiple topics and that the same terms may be used in different epistemic contexts (i. e. for different topics) with different frequencies. This approach largely avoids the problem of document misclassification for those documents sharing themes that belong to different clusters; it also allows to account for term polysemy (DiMaggio, Nag and Blei, 2013).

The method assumes that a probabilistic model describes how words are (randomly) generated in documents and fits the chosen model to the data (Blei, Nag and Jordan, 2003; Blei, 2012). In the simplest model – the *Latent Dirichlet Allocation* (LDA) model - words are generated by the mean of a latent (hidden) variable which says, for each word of a given document, which topic is developed at this particular point of the document. Once the topic is known, the term observed is generated by the term frequencies for this topic, which do not depend on the document. Therefore, each document is characterised by its topic frequencies and each topic is characterised by its term frequencies. Both distributions (i.e. document:topic and topic:term matrices) are estimated while fitting the model to the data. In other words, a topic model fits a given document:term matrix to a product of a document:topic matrix with topic:term matrix, thus reducing the document space dimension to the number of topics. A distance is chosen between topics and used for their mapping. Documents are not directly represented on such a map. They can be arranged in overlapping clusters where the weight of topics is above a chosen threshold.

[2] http://www.europarl.europa.eu/plenary/en/parliamentary-questions.html#sidesForm

An important technical choice in topic modelling is the number of topics used to describe the issue. Given the complexity of obesity, we first fitted a model with 200 topics. However, this fine-grained description seems too complicated to read. To get a bird eye view description of the variety of research objects and methods and of their respective weights in the corpus, it was then necessary to group these 200 topics into clusters. Instead of this two step procedure, such an overall description could as well be obtained with a lower number of topics. Thus, we ran a model with 20 topics, which provides an interpretable description for a first level analysis of the corpus.

We fit the LDA model with the Mallet package (McCallum, 2002) to various corpora and we analyse and display the results with R scripts and interactive visualisation provided by the LDAvis program (Sievert and Shirley, 2014). On a LDAvis figure, topics are close if they have similar term distribution. Topic dissimilarity is calculated with Jensen-Shannon divergence between the distributions, represented in a two dimensional space with the first two dimensions from a principal coordinate analysis.

For the largest science corpus (i.e. corpus C), after removing standard stop words and terms appearing in less than 100 documents, we get a list of 12,428 terms. We then fit the document:term matrix to the 20 topic LDA model. Resulting topics have weights varying from 8.2% of the corpus for topic S9 (denoted 9 on the figure) to 2.3% for topic S12. Labels are obtained by selecting a few meaningful terms among the 20 most frequent (or relevant) terms displayed by LDAvis (relevance parameter $\lambda = 0.6$).

## 3. Science supply side: mapping the research topics

We apply a topic model to the largest science corpus (i.e. corpus C) with a choice of 20 topics, as described in the Method subsection above. Fig. 2 shows meaningful terms among the most frequent ones, the relative size and the relative semantic proximity of these topics. The figure suggests four semantic clusters covering 18 topics and 2 more isolated topics:

- A cluster dealing with general biology methods on the left hand side: topics S8, S10, S15, S3;
- Another cluster related with various diseases, at the top of the map: topics S11, S17, S16 and S20;
- On the right hand side a cluster of 8 topics gathering issues on health risks related with obesity (S1, S9, S12) and on medical treatments (S13, S5). These topics also share common vocabulary with topic S4 about obesity observation and measurement and with two other topics on diet (S7) and physical exercise (S2). These latter two topics are about lifestyle, defined by Lalonde (1974) as “aggregation of decisions by individuals which affect their health and over which they more or less have control”.

- Two isolated topics deal with children activities (S18) and health care and social environment (S14). Topic S18, focused on obesity of young people, is dealing with lifestyle and with social environment and this explains its position between topics S2 and S14;

- A last cluster of two topics with a small topic on genetics S6 (genetic causes of obesity) and a large topic S19 labelled as 'Recent studies'.

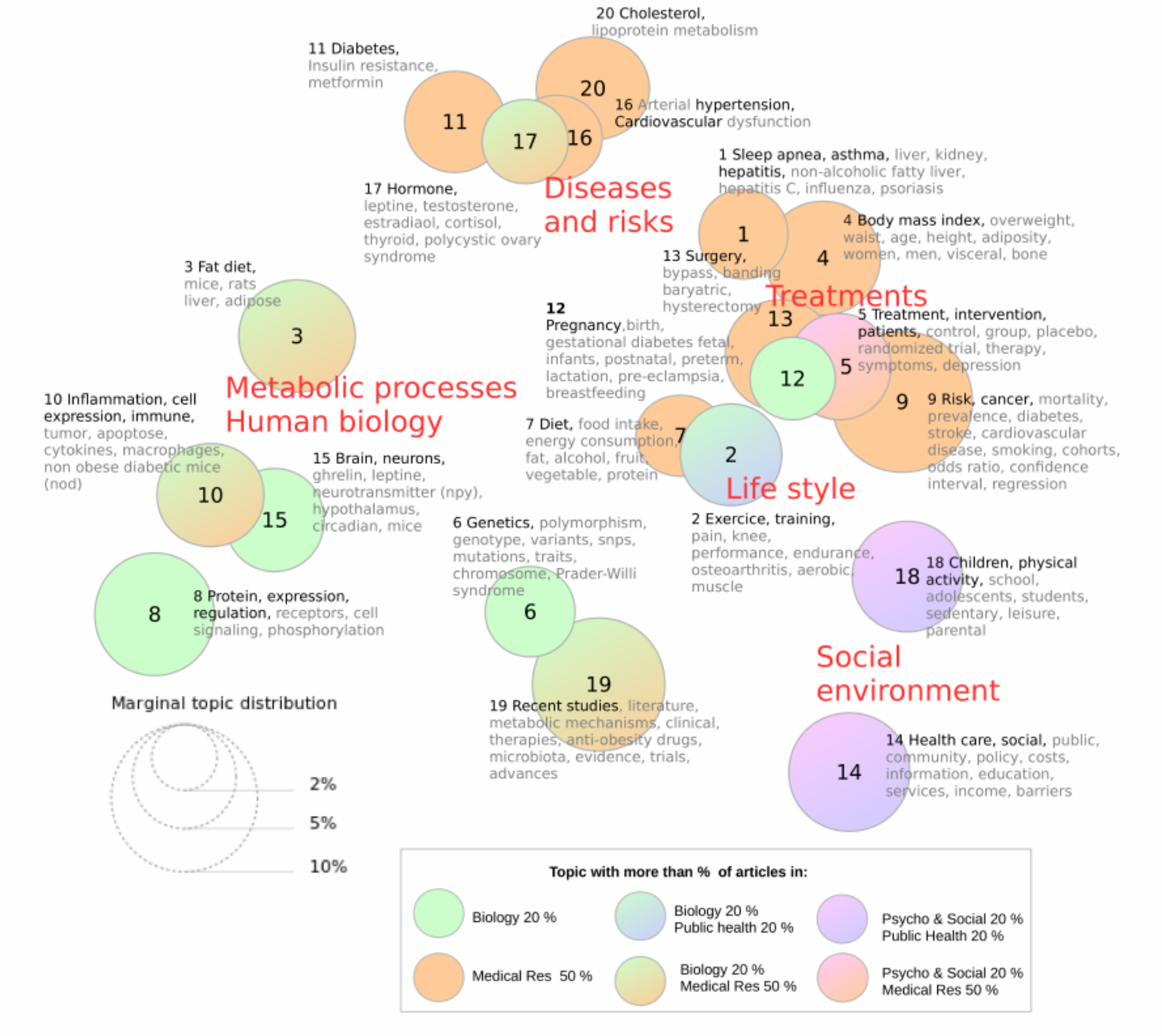

**Fig. 2. Map of the 20 topics of the science corpus fitted with the LDA model and visualized with LDAvis[3].** Colours correspond to the disciplinary characteristics of the topics as shown in Fig. 3 with different colours for topics with a share larger than 50% in medical research, and larger than 20% in other domains.

[3] https://cran.r-project.org/web/packages/LDAvis/LDAvis.pdf

Topic S19 deserves a special mention since it is difficult to interpret. To understand a topic, besides analysing its most relevant terms, the titles of its most characteristic documents (i.e. the documents with the highest weight within this topic) also provide useful information. This set contains articles about specific biologic environments such as gut flora, nutritional genetic, epigenetic and about innovative therapies or new drugs. We labelled this topic as 'Recent studies' to reflect its innovative nature and that it grew significantly over the time period.

Another way to characterize topics, different from their semantic proximity, is to consider the aim of the research on each topic. These aims could be:

- understanding biological processes related with obesity (biological causes),

- describing health risks and diseases related with obesity,

- developing medical treatments of obesity and related diseases,

- providing evidence on the risks generated by individual behaviour (life style) and social environment and support public policies to prevent obesity.

These different aims are found in the topics located on Fig. 2 from left to right in counter-clockwise direction and the corresponding zones indicated with red titles. This user oriented point of view provides a more fuzzy classification of the topics, which does not match exactly with the semantic clusters. However, such an interpretation is relevant when the objective is to highlight research driven by societal needs.

It is worth exploring the relationship between the classification of obesity based on topics - (which are only based on the terms used by scientists in their articles) - with a classification based on journals disciplines. Assigning fractional counts to topics in each document, it is possible to build a disciplinary profile for each topic. Fig. 3 shows the disciplinary profile of topics across Biology, Medical Research, Public Health, Social and Psychology (see Appendix A for details on how journals articles are assigned to these disciplinary groups).

Medical research is the most important group, with a minimum weight of 30% in each topic, representing in total 55% of the whole corpus. Observing the disciplines at a lower scale of aggregation (not shown) reveals a highest contribution of *Nutrition & Endocrinology* in topics S7 (Diet) and topic S11 (Diabetes) and of *Surgery & Gastrology & Urology* in topic S13 (Baryatric surgery). Unsurprisingly, the topics with the highest share in biology (*Agroofood, Bioengineering, Biochemistry, Reproduction & Biological development, Biotechnology & Genetics* and *Microbiology & Immunology*) are the topics on human metabolism and biology. Social sciences and Public Health publications represent about a half of the two topics on lifestyle and social environment.

In summary, the main disciplines of each topic are consistent with their first interpretation, but it is worth noting that all topics are addressed by several disciplines. This shows that the analysis of the corpus from the perspective of topics is not equivalent to its analysis with conventional disciplinary categories. This analysis confirms that obesity research is multidisciplinary (i.e. diverse disciplines are involved), but we have not analysed at this stage its possible interdisciplinarity (extent of disciplinary integration) but it could be examined when focusing on particular topics associated with desirable research options. Such an analysis would allow for further understanding of the research represented by topics S14, S19 and S2 for example.

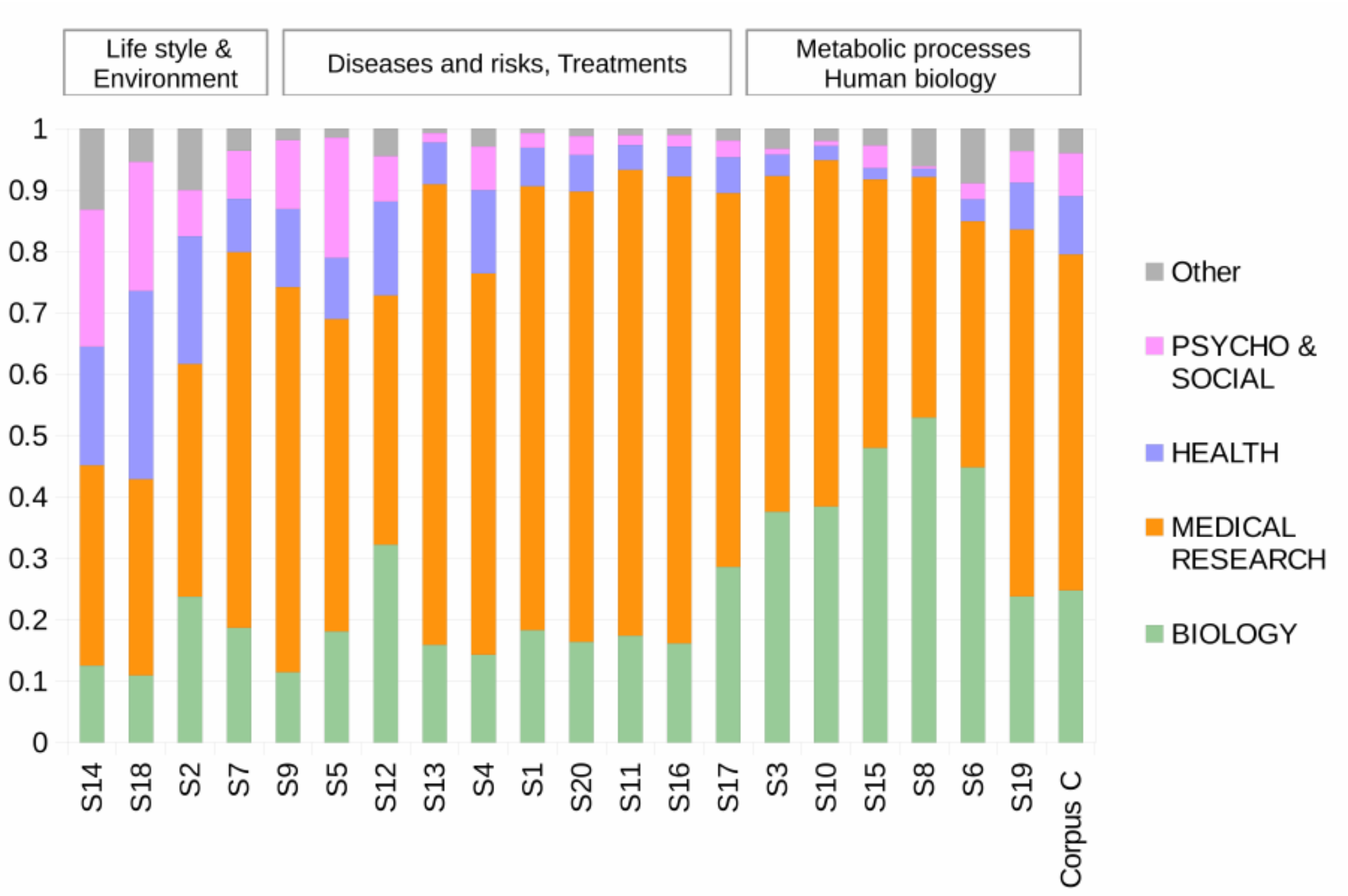

**Fig. 3. Disciplinary profiles of the 20 topics in four groups of disciplines.** Topics are ordered counter clockwise, according to their proximity on Fig. 2.

As mentioned earlier, topics do not cluster documents but describe themes which are shared in the corpus. In the present case, the distribution of the weight of the heaviest topic by document shows that only a small proportion of documents are focused on a single topic. For instance only 15% of documents have a topic with a weight greater than 0.75 and only 55% have a topic with weight greater than 0.5. Documents with a high topic weight (say larger than 0.5) could be considered as core or specialised documents for topics, but among the 20 topics, only three have cores with more than 13,000 documents (out of a whole corpus of 287,000). These are topics S8, S9, and S14 (representing respectively, 10%, 9%, and 8% of the whole set specialised documents).

This means that 45% of the documents mix topics for more than a half of their content. The intensity of these topic combinations is largely a consequence of the big size of the corpus and of the low number of topics. Topics do not provide the same type of description of the corpus as would do a classification with hard clustering constraint (i.e. not allowing overlapping clusters) but avoid misclassification of intermediary documents bridging various viewpoints or methods for instance.

Topic proximities on Fig. 2 are based on term frequencies: two topics are close if they share common terms. Notice that this is different to examining the issues which researchers see as connected, which can be explored by looking at those topics co-occurring in documents. In summary, while the topic:term matrix information used for Fig. 2 tells about semantic similarity of topics, the information contained in the document:topic matrix informs about which topics are published together. For this latter objective, we can for instance look at documents that share two topics where both of them have a weight larger than 0.25. In our corpus, 52% of documents have this property. To show which topics are co-occurring in documents, we use a network representation where edges between the nodes indicate pairs of topics sharing at least 1,500 common documents where each topic weighs more than 0.25. In order to help interpretation, on Fig. 4, we position nodes approximatively as in Fig. 2 where co-occurrence links are shown through edge thickness (and therefore not as proximity between nodes).

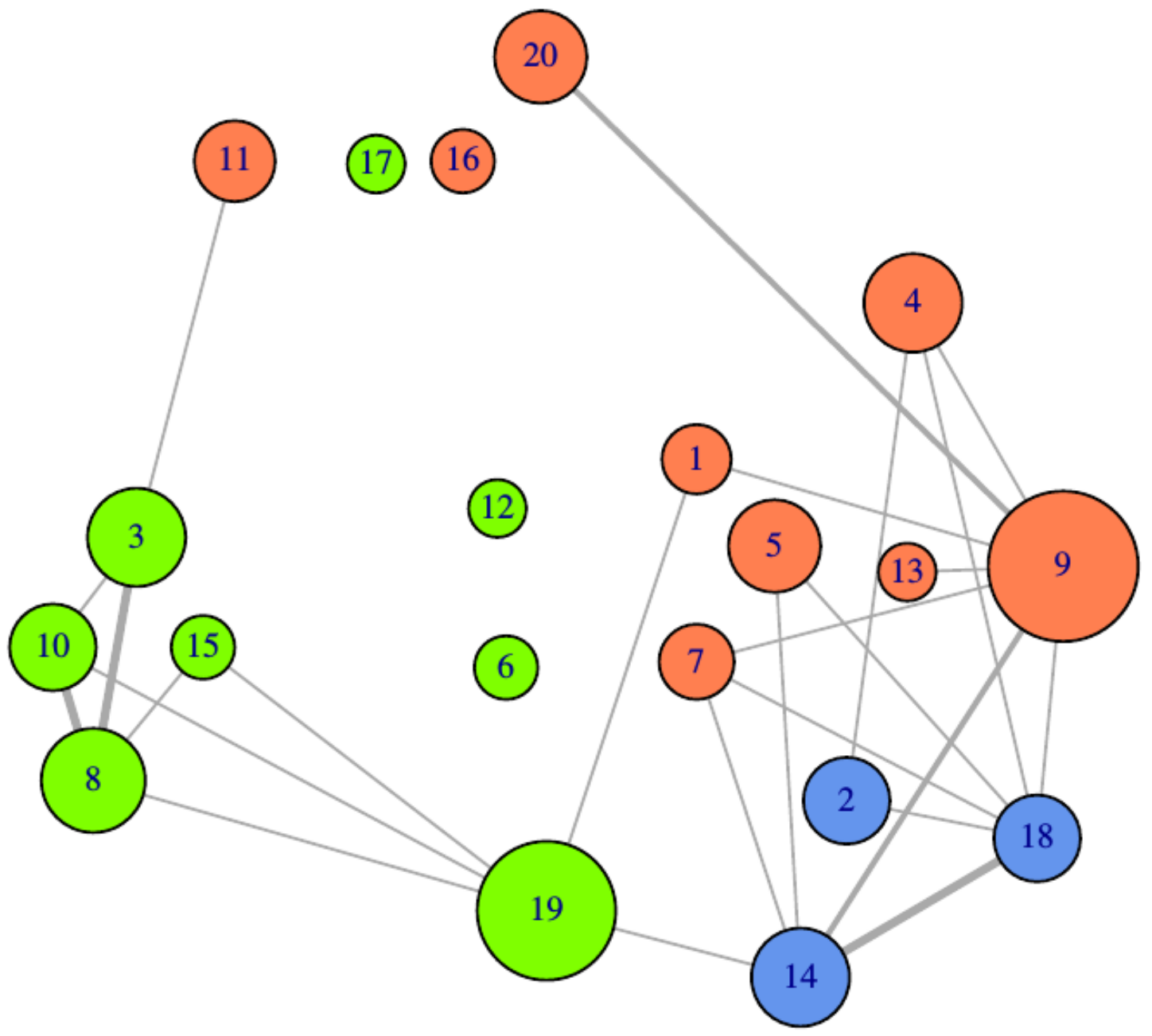

**Fig. 4. Network representation of topic co-occurrence in documents of the science corpus C.** Edges are present when the number of documents where both topics weigh more that 0.25 is larger than 1,500. Edge thickness shows three levels of topic co-occurrences: when the number of shared documents is between 1500 and 2500, between 2500 and 3500 or larger than 3500. Node size is proportional to the number of documents shared by a topic with all other topics (i.e. the degree of the node).

We observe the following patterns in Fig. 4:

- A first group of biology topics includes S8 (Cell biology), S10 (Inflammation) and S3 (Fat diet), S15 (Brain) and S19 (Recent studies). A second larger group gathers the topics on social environment and life styles and the topics about treatments. It is only linked with the first group through topic S19.

- Two topics have a large degree: S19 shares about 24,000 documents with other topics. As it connects the two groups, this suggests an interdisciplinary feature of this topic.

- Another topic has a high degree: S9, the general topic on risks, shares about 26,000 documents with other topics. Among them are topics about risks and treatments as S20 (cholesterol), S1 (apnea, hepatitis), S13 (surgery), S4 (obesity measures) and S7 (diet). But S9 also shares documents with the social environment topics S14 and S18. This topic may be understood as a transversal topic summarising the various risks related with obesity.

- Topics S14 and S18 are strongly linked together (with 3,820 shared documents). With an important part of social sciences research, and an important core (22,500 documents with a weight larger than 0.5 for one or the other topic), they are also connected to other topics as S5 (general treatment), S9 (health risk) and S7 (diet, food intake). This pattern suggests that social aspects and health care issues are a transversal issue.

- Unlike these transversal topics, topic S12 (pregnancy), S13 (surgery), S17 (hormones) and S6 (genetics) appear to be more isolated (sharing the lowest numbers of documents with all others).

In summary, the proximities associated with topic co-occurrences in documents are different from semantic proximities related with term usage and these differences would require some refined analysis. For both representations, the biology topics form a separate group which share the same basic biology vocabulary as well as common documents. These topics or documents can be considered as some basic knowledge useful for understanding biological processes involved in obesity but not tackling the obesity problem directly. They could be analysed separately, provided that topic S19, which is part of this specific analysis, is not withdrawn from the core corpus on obesity research. On the contrary, social and life style issues are intertwined with at least 4 topics in medical research and this part of the corpus suggests that interdisciplinary or holistic scientific approaches exist. If there are provided with more precise information of such features, experts and stakeholders could discuss whether such transversal approaches should be encouraged. Finally, the specificity of those topics treated in isolation as cardiac and cardiovascular or endocrine dysfunctions and diseases, as well as pregnancy obesity related problems could also be commented by experts.

*Time dynamics*

Another issue is the dynamics of the topics. As the selected corpus covers a large period of time, and because the concern about obesity increased along time, it is useful to compare topic profiles over time. Fig. 5 shows that the proportion of the two social topics S14 and S18 in the corpus increased around 40% in the whole period as well as the transversal topic of recent studies (topic S19). This is balanced by decreasing relative weights of specialised biological topics as topics S3, S8 and S15, as well as those of medical research on diabetes (S11) and endocrinology (S17). We will see later that three of the increasing topics are those mostly related to societal demands as captured through the MEPs questions.

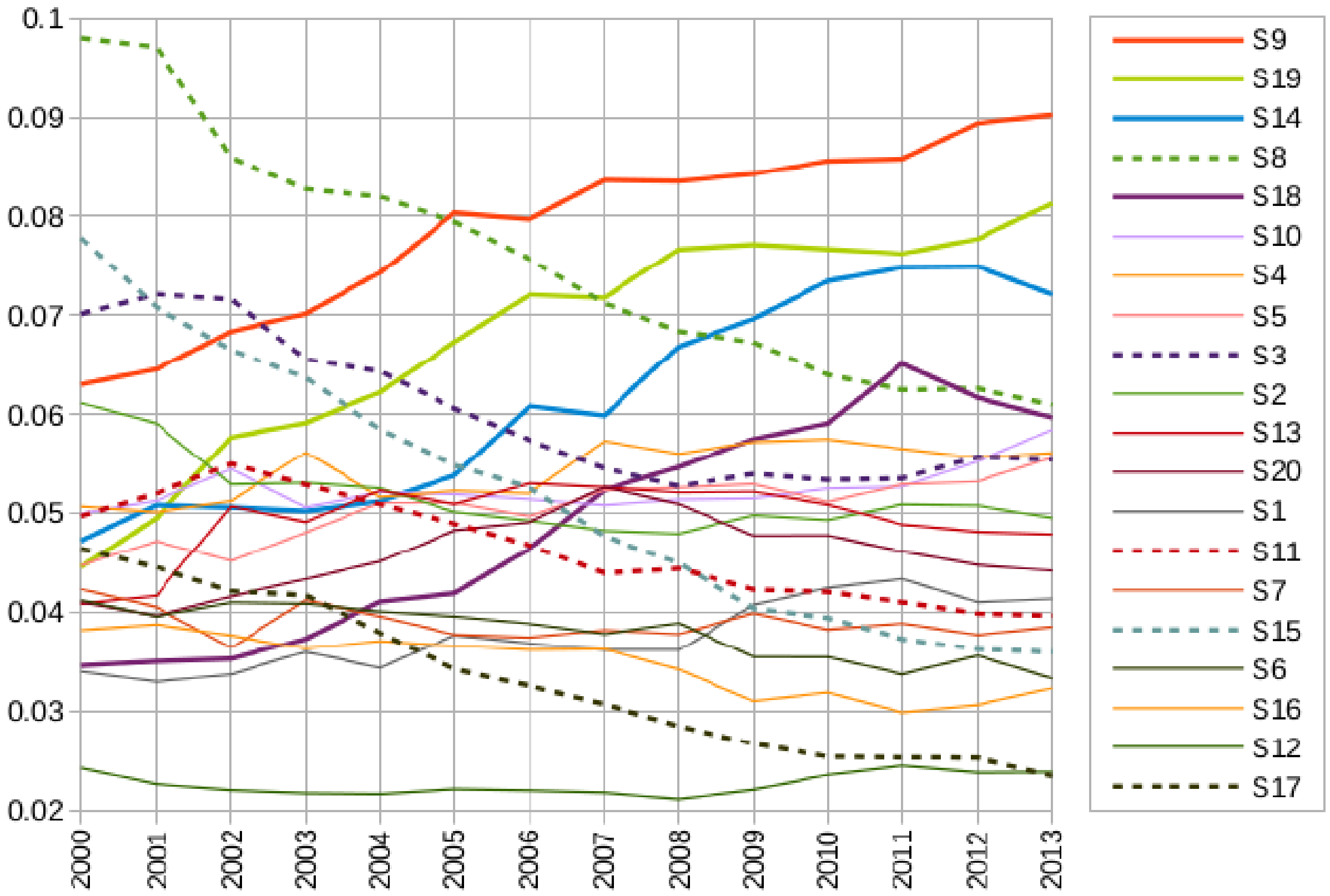

**Fig. 5. Variation of topic weights over time.** Solid thick lines show the topics that increase over time, while thick dotted lines show topics that decrease in the same period.

Our large scale analysis of a comprehensively delineated corpus can be compared with a smaller scale analysis (3,545 publications) carried out by Nicastro et al. (2016) who summarize obesity-related research funded by the US National Heart, Lung and Blood Institute (NHLBI) 1983-2013 using a bibliographic coupling network (Fig. 2, page 1359). Our biology and medical research topics can be broadly related with 17 of their communities while some aspects of our life style and social environment topics are found in their two clusters named *Correlates & consequences:obesity* and *Correlates & consequences:weight change*. On a much smaller corpus focused on the research funded by an NIH institute, they get a more fine-grained classification of the biological and medical topics, but non medical issues are less represented than in our corpus. This suggests that these latter aspects deserve a more detailed analysis in order to illuminate research in other disciplines which have a lower coverage in PubMed or WoS databases.

## 4. Societal demands side: European parliament questions

The same topic modelling method is used to map an instance of societal demands using 222 questions in the European parliament for 2009-2014. After various tests, we chose to adjust a rather large number of topics (30) to this small corpus because of the variety of issues covered. Even so, it is not easy to give a title to each topic. Rather than the most relevant words of each topic, we use the titles of its characteristic documents - i.e. with topic weight greater than 0.85 - and summarize them into a topic title. This method is relevant because, unlike the science corpus, the documents of this corpus do little mixing of topics : clusters of characteristic documents cover 86 % of the corpus.

Topics can be classified into four groups as shown in Table 2 and Fig. 6. Two groups deal with the causes of obesity that the MEPs cite before asking the Commission what measures will be taken: 11 topics about dangerous substances as endocrine disruptors, unhealthy food and diet (pink bubbles on the right hand side of the map), and 6 topics about adverse effects of food industry, advertisement strategy, and market deregulation (blue bubbles in the centre). These two groups represent about a half of the corpus. One topic (topic E8) contains parliamentary questions that underline social factors such as poverty and malnutrition.

Another group of 7 topics, at the left hand side of the map, contains propositions of actions to prevent obesity (green bubbles). This group weights about 30% of the whole set.

Finally, 10% of the issues reported by MEPs deal with health consequences of obesity such as diabetes, cancer, cardiovascular problems and impact of obesity during pregnancy (yellow bubbles). A particular topic about diabetes epidemic corresponds to a single question that was asked by four different MEPs. This produces an isolated topic (top left of Fig. 6) associated to the specific distribution of terms of this question.

*Table 2:* Titles of European parliament topics. Topics are listed by cluster and according to their proximity on Fig. 6, in a counter clockwise direction.

| Cluster | Topic # | Title |
|---|---|---|
| **Hazardous substances and diet** *28%* | E20 | fructose acceptance and regulation, dha (omega-3 fatty acid) , reference values for children intake |
| | E18 | artificial sweeteners, nutrient profile assessment |
| | E7 | palm oil, endocrine disruptor, meat consumption, food production |
| | E6 | hydrogenated fat, energy drinks, food labelling, processed food, light pollution |
| | E27 | bisphenol ban |
| | E11 | livestock production, bisphenol, cigarette packaging |
| | E16 | endocrine disruptor, anti-obesity drugs, discrimination of overseas countries |
| | E17 | salt consumption, dietary legislation, pesticide use, kidney failure, training doctors, |
| | E28 | hepatitis, bisphenol in food containers |
| | E5 | appetite suppressant drugs, poor eating habits of young people |
| | E14 | junk food, fizzy drinks, obesity in Mediterranean and worldwide |
| *Intermediate topics* | E13 | protecting young athletes, combating children obesity, bad quality meat, diabetes drugs |
| | E10 | drinking water, prevention with breast milk, red label products |
| **Market and food industry adverse effects** *24%* | E22 | food advertising for children, product formulation, strategy effectiveness |
| | E26 | food industry strategy, food costs, research on nutrition, good practices |
| | E2 | food advertising at children |
| | E9 | agriculture subsidies, market deregulation, obese people penalized |
| | E1 | deregulation of sugar market, tax on sweets |
| | E4 | junk food tax, stevia |
| *Intermediate topic* | E8 | malnutrition, food safety, Mediterranean diet, poverty |
| **Prevention and remedies** *31%* | E15 | research protocol, prevention, cost reduction |
| | E29 | prevention programmes, assessment of strategy, healthy lifestyle promotion |
| | E12 | sport and physical activity, fruit at school scheme |
| | E21 | education of young people, data on obesity, food taxation policy |
| | E19 | fruit & vegetable at school, women malnutrition, food labeling |
| | E25 | school food, children friendly labelling of food |
| | E24 | fast food, healthy eating education |
| **Related diseases** *10%* | E3 | diabetes, cancer, risk factors |
| | E23 | heart attack and hearing loss risks, gestational and juvenile diabetes, research |
| | E30 | diabetes epidemics |

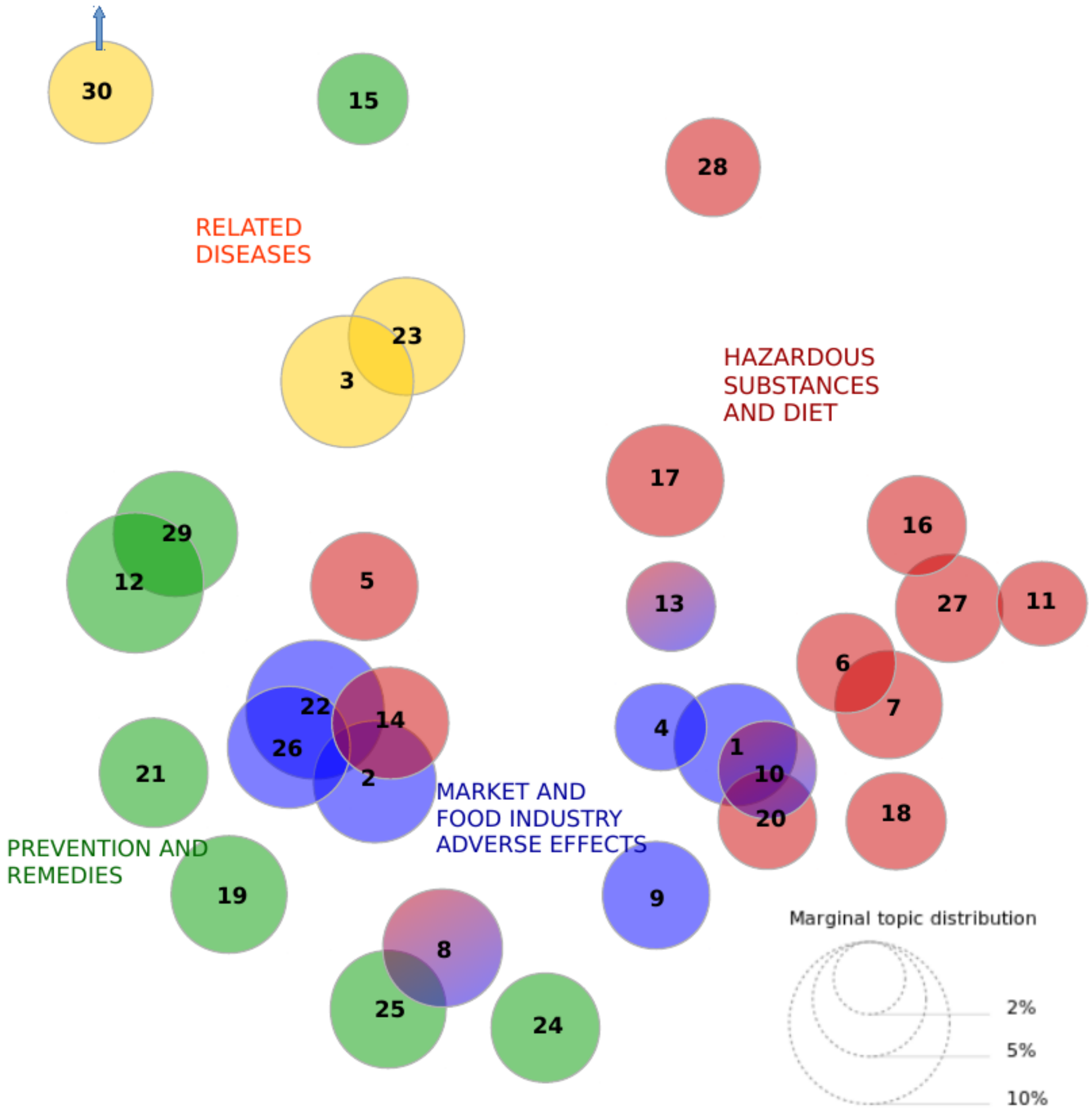

**Fig. 6. Topic map of the corpus of the questions from members of the European parliament (MEPs) to the European Commission** (222 questions, October 2009 - February 2015).

Before carrying out an in depth comparison of the societal topics with the science topics, it is important to note that the issue of dangerous substances is not visible in the science corpus at the chosen observation scale. The same remark holds for most economic (industry and regulation) aspects. Recommendations for political actions (except about physical activity) are also missing in the science corpus which means that there is no evidence that these recommendations are sufficiently backed by scientific knowledge. On the contrary, topics about related diseases are widely treated in the science corpus. In the next section, we try to quantify these observations of overlap between scientific and political discourses.

## 5. Exploring alignment between societal demands and research supply

Having summarized science supply and (one of the expressions of) social demands with topic modelling, we now compare the two corpora through their topics. We should stress that this comparison is exploratory and should be interpreted with caution because the two corpora are made from two different discourses, namely scientific vs. politics/policy discourses.

Each topic is portrayed as a distribution on a concatenated vocabulary, which is composed of the terms from either corpus. The science vocabulary (12,428 terms) and the European parliament vocabulary (4,466 terms) share 2,976 terms. This allows us to calculate a distance between each pair (Science topic, EU topic). We use again Jensen-Shannon divergence as the distance between two topics and the averages of these distances on either science topics or EU topics. These averages are shown as left and bottom margins of the table shown in Fig. 7 and there are also represented on Fig. 8, where topics with a thick border are those which have some echo in the other corpus. Edges across the two maps represent the 23 closest pairs of (Science topic, EU topic).

On the science side, the social environment topics, S14 and S18 and topic S19 (on innovative treatments based on recent studies) are the closest to the European parliament corpus. Topic S7 (Diet) and topic S9 (Epidemiology of risks) – also partially meet European parliament concerns. On the contrary, fundamental biology topics S6, S10, S15 and S17, and topics about specific diseases S1, S11, S16 and S20 are not very much related to European parliament concerns.

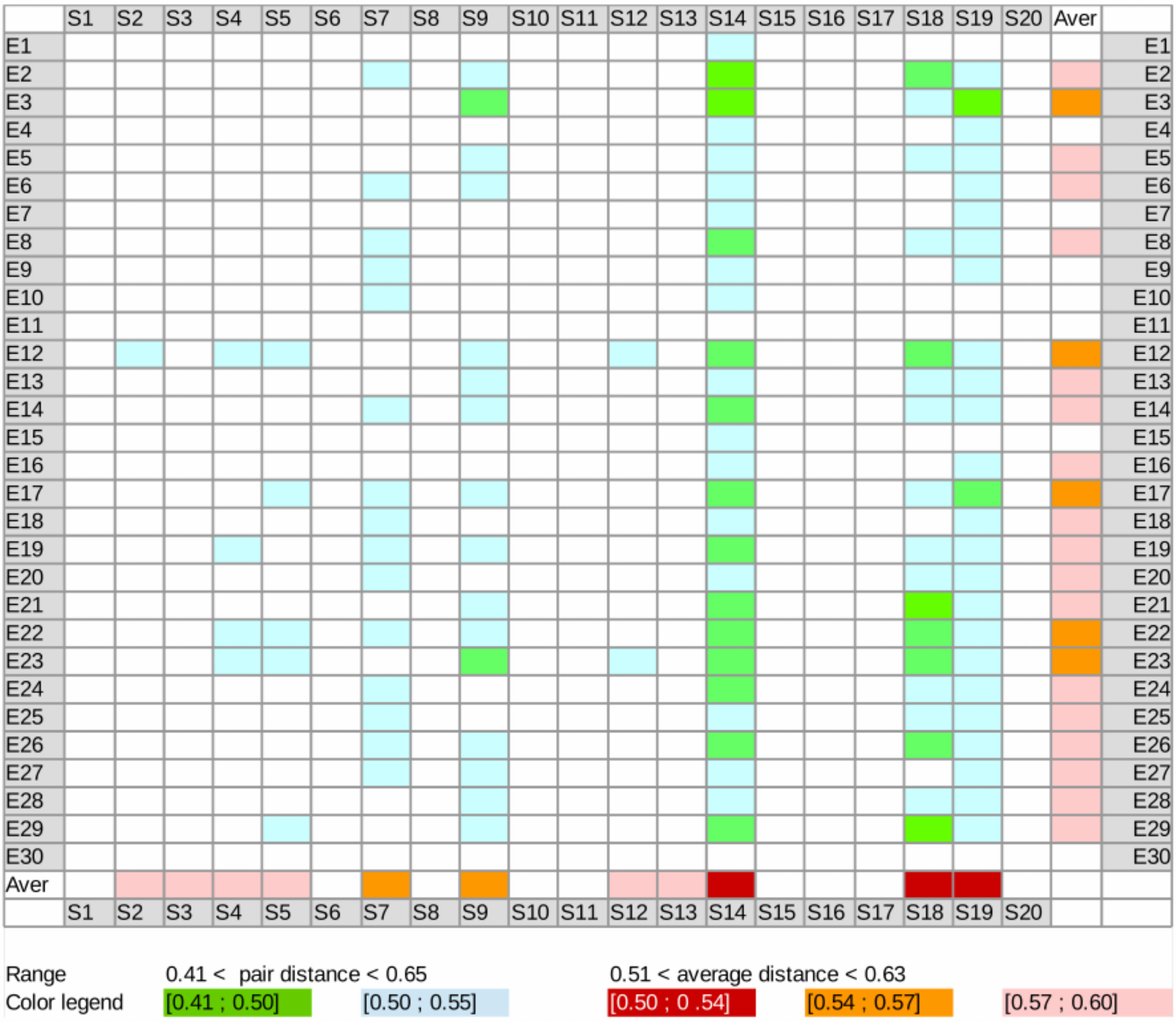

**Fig. 7. Distances between science topics (based on scientific publications) and societal EU topics (based on European parliament questions).**

From the parliamentary side, a few topics seem to be partly ignored by the scientific community. This is the case, for example, of some economic policy topics about sugar market deregulation, food taxes or agriculture subsidies (topics E1, E4, E9). So is topic E15 on health economics – prevention, cost reduction, research protocols. On the contrary, the food advertising issue (E22) has some correspondence on the science side. This is also the case for the issues about diseases related to obesity (E23, E3) and for sport and physical activity (topic E12).

We also note that four of the topics connected to societal demands, namely S14, S18, S19 and S9, had a faster publication growth from 2000 to 2013 (Fig 5).

However, it would be dangerous to interpret the concordance of the scientific maps with MEP maps as an impact of the policy on scientific agendas. These two corpora are constructed with different type of documents - scientific articles vs. political questions - and therefore lack of similarity can be due to a variety of reasons related to the type of semantics used. Where there is similarity, however, it suggests that the research has a more direct relation to policy. The detailed examination of the topics with highest similarity (e.g. those related to public health and social determinants) supports this view. However, we should stress that scientific topics without direct similarity with MEP topics may be relevant to policy. They may be dissimilar simply because they use a different language or because they contribute indirectly to another topic that has more immediate policy relevance.

Consequently, is not possible to infer some causality between the two phenomena. We should also remember that parliamentary questions is just one instance of traces to capture societal needs. It is insufficient data source and thus only a partial indicator of societal demands. But other type data from public discourse (e.g. from Twitter) will also have the problem of belonging to a discourse different from the scientific – and hence the comparison will be problematic. We believe that expert interpretation will always be needed to check if the linkages captured are meaningful.

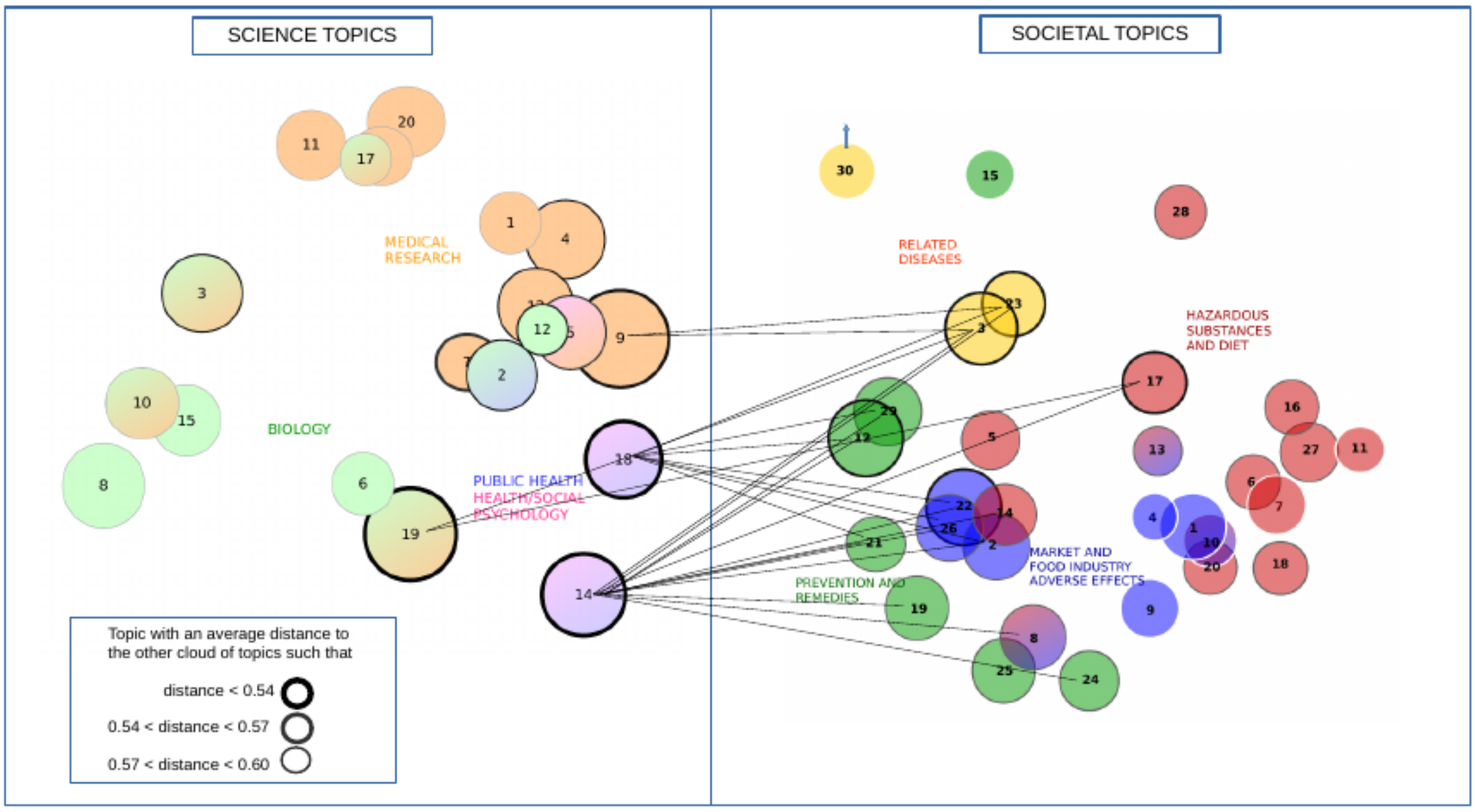

**Fig. 8. Vis-a-vis of the two topic maps.** Dark borders show individual topics that are closer to the whole topic set of the other side. Lines between the two maps show topic pairs with a distance lower than 0.5.

## 6. Zooming on social issues

The choice of a large corpus to represent the supply side ensures that no significant research related with obesity has been discarded. However, this large scope together with the coarse grain classification of topics we have adopted means that it is unclear how the relevant issues in policy are supported by scientific research. Moreover, social issues related to obesity are likely to be less represented or less visible in the scientific literature given that the databases we used have a far more comprehensive coverage of the biomedical sciences and do not include books.

Given that the previous section provides evidence on the relevance of social environment topics (S14 and S18) for policy, we develop a more fine-grained approach for these topics, as to test if a description more directly related to policy choices can be achieved. We define a restricted corpus by selecting those documents that have a weight greater than 0.25 for either topics S14 and S18. As these documents share many documents with topics S19, S7, S5, S9 and S4 (Fig. 4), this corpus will also contain documents addressing social issues jointly with medical or biological approaches. Interesting, among these topics there are three topics (S19, S7 and S9) that are, as S14 and S18, among the most connected with parliamentary questions. This selection process therefore gathers a large part of the scientific literature that could support policy options. This choice leads to a restricted corpus SC of 46,507 documents, corresponding to 16.78% of corpus C.

Again, we analyse the corpus with a topic model. We fitted 10 topics to the data (referred to as U1 to U10) and we report a few keywords selected from the 30 most relevant terms on a LDAvis map (Fig. 9). We also examine the proportion of WoS categories in these topics (Fig. 10) a well as the WoS categories of characteristic papers per topic, namely of those papers with topic weight greater than 0.9.

Fig. 9 provides a refined description of the two selected topics and of their semantic neighbourhood.

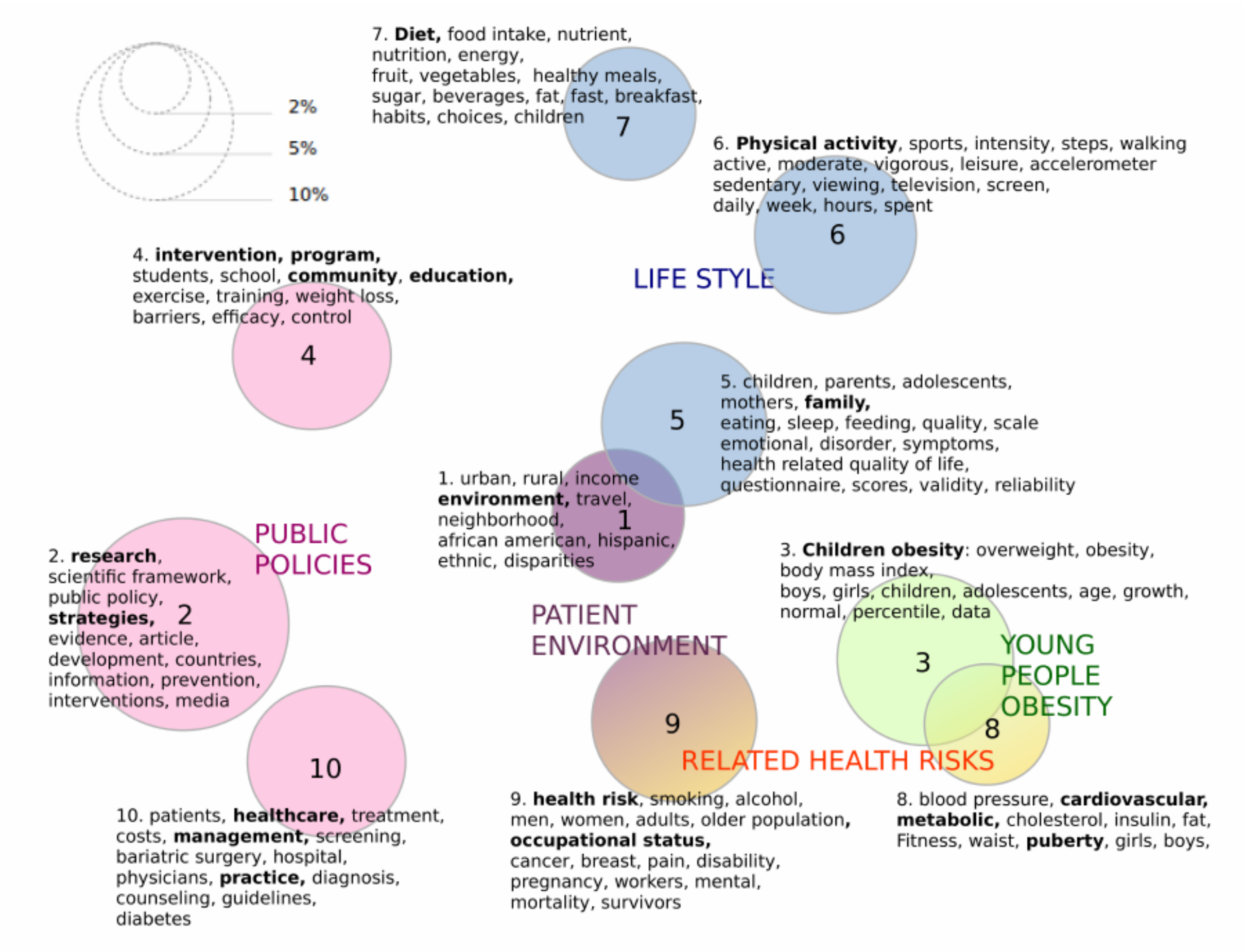

**Fig. 9. Map of 10 topics of a social environment corpus.** The corpus is the subset of 46,507 documents of the science corpus where topics S14 or S18 weigh more than 0.25.

The cores of the previous social topics are now split into three topics on public policies and another topic about the impact of the social environment on obesity:

- Topic U10 addresses issues on health management, health costs and clinical practice.
- Topic U2 gathers research work on public research agendas. This is the only topic with characteristic documents (with topic weight larger than 0.9) published in the following social sciences WoS categories: *Sociology (6), Social Sciences Biomedical (8.4), Communication (20.3), Economics (5.7), Management (5.8), Political Science* (5.5) and *Ethics (6.3) as well as Food Science & Technology (7.5)* and it has the highest number in *Health Policy & Services (11.7) and Business (28.5).*

- Topic U4 deals with education and intervention programs in communities and schools. An inspection of the titles of these documents (Appendix B) confirm this interpretation. This topic has a total weight equivalent to 4,476 articles but it is widely scattered across documents with only 16 characteristic documents.

- Topic U1 deals with local geographic, ethnic and social factors of obesity (36% of articles in *Public, Environmental & Occupational Health*). The titles of the 26 characteristic papers in Appendix C illustrate this interpretation.

Linked with these policy and social topics, three topics bring together the issues about life style that were already identified with previous topics S7 (diet), S2 (exercise) and partly in S18 (children and physical activity)

- Topic U7 is about diet, similar to previous topic S7. A large percentage of the articles are published in category *Nutrition & Dietetics* (38%) (Fig 10).

- U6 deals with physical activity and its measures (time devoted, intensity...) as previous S2. It has the highest proportion of articles in *Sport sciences* and some characteristic papers on this topic are published in *Physiology* journals (18 papers).

- U5 deals with the impact of family life style on obesity, and also with its links with psychological disorders. Unsurprisingly, this topic has the highest percentage of articles in psychology (11%) and there are characteristic papers in the three categories *Psychology Clinical*, *Psychology Biomedical* and *Behavioral Sciences* (48 characteristic papers).

A focus on children obesity is finally identified in topics U3 and U8. Topic U3 deals with the description of obesity regarding its biological and human aspects, with a specific focus on young populations, consistent with the highest percentage in *Pediatrics* of this topic. This is also the only topic with characteristic papers in *Anthropology* (15 documents). Health risks factors are gathered in topics U8 and U9 (respectively for young people and adults). These three topics are partly coming from previous topic S9, which is strongly linked with S14 as shown in the network representation on Fig. 4.

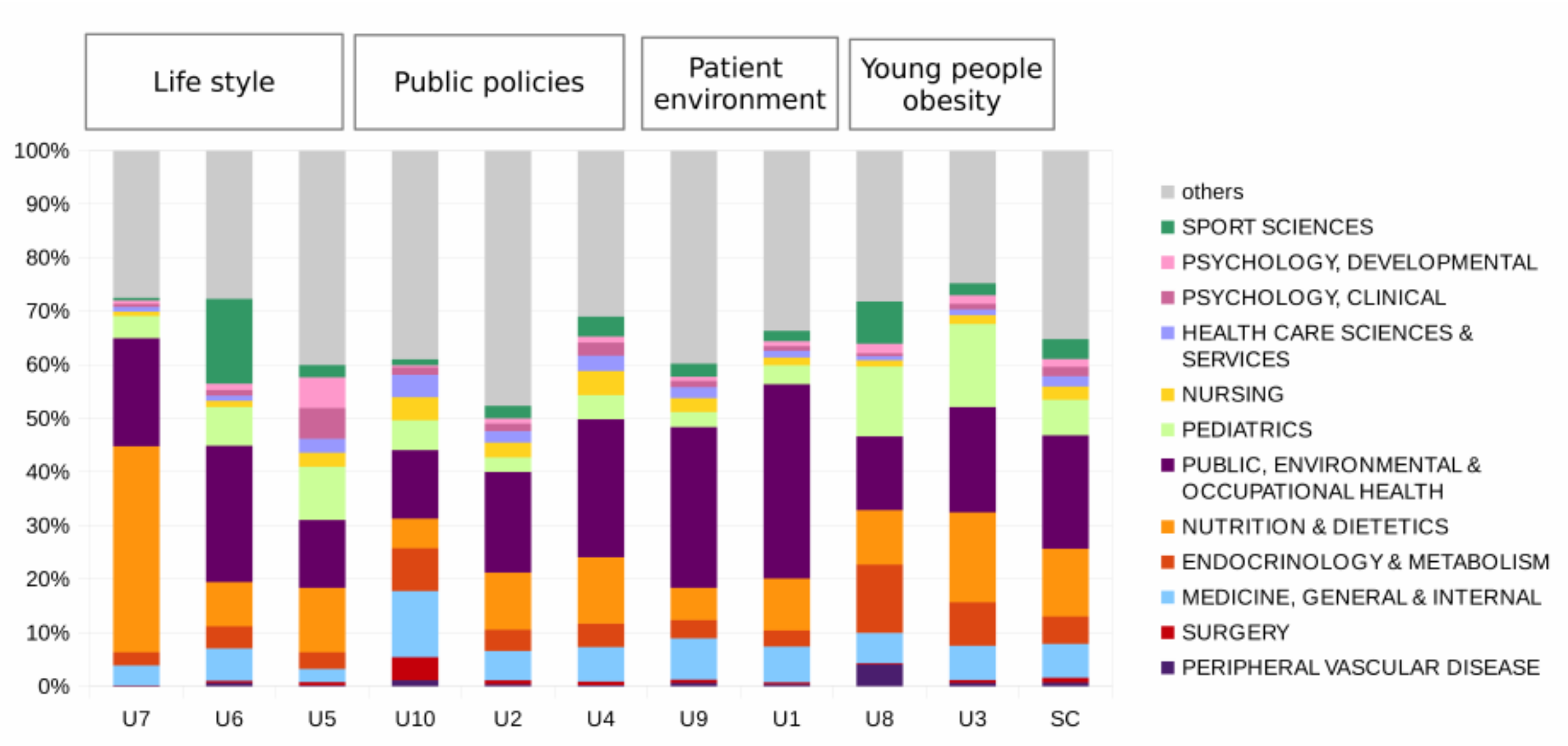

**Fig. 10.** Distribution of WoS categories in the 10 topics and in the whole corpus SC. Only categories with a percentage greater than 4% in at least one topic are represented.

This restricted corpus reveals research work on socio-political issues such as as family environment, living environment, community interventions, which have been stressed as particularly relevant to address obesity by experts (Popkin, 2007, 2009, Christine Cherbut[4]) and governments (Government Office of Science, UK, 2007; Basdevant, 2013). Life style topics and public policies in particular are much more explicit in this detailed model than in the first analysis. Such a fined-grained approach (about 4,500 documents per topic) seems more appropriate to capture experts views and support stakeholders' deliberation.

As the 20-topics model is adapted to provide an overall view of the large corpus, a description on a lower scale tested in this section seems relevant to better understand the various scientific options and research works. This smaller scale analysis should be achieved for other topics as for instance topic S19 and its biology research neighbourhood, which is not well understood at this stage of the analysis. This suggests that, in a further step, a model with a number of topics between 50 and 100 might be worth fitting on the whole corpus. However, some other detailed topics may not deserve more attention when focusing on obesity. Specific medical research topics such those on diabetes, cardiovascular or respiratory diseases and treatments are less relevant with the aim of this study and, due to our inclusive delineation method which ensures a good recall, these side topics would also be described more precisely. If a better precision on obesity research is wanted, characteristic documents of such topics could then be discarded from the corpus. The delineation process could therefore be improved in a satisfactory way.

[4] Christine Cherbut, INRA Scientific Director for *Food, Nutrition and Bioeconomy*, Personal communication (January 2015)

## 7. Robustness of mapping techniques

The technical purpose of this article was to experiment with topic models as a method to analyse a corpus of scientific publications and to compare it to a body of non academic documents. Let us now discuss its robustness.

We acknowledge that the study follows adhoc choices in parameter setting that makes the study potentially idiosyncratic (as it is the case of most bibliometric research). First, the use of PubMed and Web of Science may have biased the sample towards biomedical approaches, potentially excluding relevant research from the social sciences (e.g. policy related) and low and middle income countries (Rafols, Ciarli and Chavarro, 2015). Second, we used a comprehensive corpus in order to include obesity research as well as obesity-related research – that later in deliberations some stakeholders my decide (or not) to partly remove (e.g. some research on diabetes). The threshold of inclusion (10% of obesity papers) is indeed questionable, but the choice of a much higher threshold (50%) did not imply a major change in relative coverage across disciplines – hence it might be that the overall map is rather stable, a hypothesis that would need further testing.

Third, topic modelling is sensitive to parameter-setting, not the least the arbitrary choice of number of topics. As explained above, the choice of 20 topics was pragmatically made as offering a map that was readable and provided topics that we could interpret when comparing the terms of a topic and the contents of specific publications. We are aware of a critique to the contents of topic model and agree that it is an issue deserving further research (Leydesdorff and Nerghes, 2017).

One way forward to test the technical choices, would be to carry out comparison with other topic extraction methods, as carried out recently by Velden et al. (2017), given the lack of a gold standard.

In our approach, we have used topic modelling rather than a simple term frequency analysis as provided by various software as VOSviewer text mining facility (Van Eck and Waltman, 2011) or CorTexT[5] for instance. For comparison, we also performed a textual analysis of the science corpus with VOSviewer, selecting the 782 most relevant terms among terms appearing in at least 1,000 documents. The overall map of terms (not shown) is similar to the topic map in Fig.2. We obtained 15 clusters that can be partially related with topics, though some clusters are very small and not easily interpretable. While this large scale description is similar to that provided with 20 topics, we believe that with a finer resolution (i.e. more topics and more term clusters), the term clustering approach might not work so well, due in particular to the hard clustering of terms which does not allow clusters to share terms.

Among alternative methods, previous studies using a large corpus of medical research have suggested that citation information or hybrid methods coupling citation and textual information, would do better than using only textual information from titles and abstracts (Boyack and Klavans, 2010; Boyack, Newmann, Duhon et al. 2011). A recent study by

[5] http://www.cortext.net/

Velden et al. (2017) has compared 8 clustering methods of scientific publications (including direct citation, bibliographic coupling, semantic similarity and hybrid citation-semantic clustering), and found substantial differences in the solutions – although all shared key features.

However, we are not aware of any study comparing topic modelling with citation approaches. In order to achieve such a study, one has to compare similar outputs, which can be either clusters of documents or term distributions. Clusters of characteristic documents associated with the topics of a topic model can be compared with citation based clusters with any od the available methods as for instance through an archipelago pseudo-map obtained by reordering the clusters, where the heaviest clusters of both clustering should form a roughly diagonal zone (Zitt, 2015). Another interesting graphical representation suggested by one of the reviewer would be to show citation similarities between topics as edges on the topic map, as we did in Fig. 4 with topic co-occurrences. Instead of comparing clusters of documents, one can compare term distributions in both approaches. For each citation based cluster of documents, the overall distribution of the terms in their titles and abstracts (selecting the same vocabulary as the fitted topic model) provides pseudo-topics which distances to authentic topics can be examined and represented in a grid similar to Fig. 7, or with a two-mode network graphical representation. These are some of the possible options to explore this issue, which would deserve further work.

Finally, regarding the mapping of societal demand, we acknowledge that the selection of a suitable corpus is far from being rigorously addressed. In this paper, we only used European parliament questions as an example to explore how topic model can compare supply and demand represented as two separate corpora. MEPs questions are useful in political science analysis, but the selected set is too narrow for obtaining a comprehensive view of policy issues, given that Parliament members is a particular group of people, embedded in political networks and probably fairly aware of EC health or scientific policy orientations. Using web information as in the Tweetoscope tool (Chavalarias, Castillo and Panahi, 2015) or media information (Cointet, Cornilleau, Villard et al. 2011; Di Maggio, Nag and Blei, 2013) could also be considered as a representation of the social demand. We do not believe that parliamentary questions, or any type of policy documents, will be a robust way of mapping societal needs. But it is an entry point that may help get started the deliberations on topics or issues that are perceived as important in social settings that may not be reflected in scientific priorities.

## 8. Discussion and conclusions

Science policy is increasingly in need of methodologies that help manage research associated with grand challenges or societal problems. These methodologies should not focus on assessing the perceived quality of research, but on whether the knowledge produced serves the mission of helping address the grand challenges (Sarewitz and Pielke, 2007; Stirling, 2015). In this article, we have presented a method that aims to help in doing so by comparing semantic analysis of publications (a representation of science supply) and policy documents (an instance of societal demands). The method is at an early stage and cannot claim robustness. Thus we do not recommend this method as a stand-alone tool, but we hope is that it may be useful for informing expert deliberations or policy discussions on the alignment between science supply and societal demands for a given grand challenge.

The development of the method had to overcome three hurdles: i) the delineation of a broad concept such as obesity; ii) the mapping of a scientific and a policy corpora; iii) the comparison between these science and societal maps. For the delineation of the scientific corpus, we used an inclusive choice yielding a large corpus in order to ensure a comprehensive recall. Such inclusiveness is important since the mapping should allow for later deliberations about what epistemic areas are relevant or not to the challenge. The mapping of documents has been carried out using topic modelling, a method that relies on text (not citations) and can thus be used both for scientific and policy (societal) documents. Topic characterisation and interpretation based on term list has been found to be sometimes ambiguous and challenging. The comparison between the science and societal maps has been conducted via the graphical representation (Fig. 8) and semantic similarity (Fig. 7), and have allowed to explore which of societal issues or social demands are directly addressed (or not) in scientific publications.

We have conducted a case study on obesity, in which the mapping of science supply (Fig. 2) shows five main areas: biology and metabolic processes, studies on health risk and diseases, treatments, lifestyles and social environment. Scientific publications related to obesity are mainly drawn from medical disciplines and biology (making between 70% to 90% of the publications in 17 of the 20 topics). However, the topics that are growing faster in relative terms are those related to social environment and which include the higher proportion of social sciences and public health articles. The map of policy documents (Table 2 and Fig. 6) shows four areas: hazardous chemicals and food, related diseases, food industry and adverse effect, prevention and remedies. Unsurprisingly, the comparison between the science and policy maps (in Figs. 7 and 8) suggests that research related to social environment (social sciences and public health) and research on epidemiology or treatments is more related to policy concerns.

Nevertheless, neither the science or policy maps provide a classification of knowledge on obesity that can be easily related to studies of public health or policy interventions, for example as described by Millstone et al. (2006), the UK Government Office for Science (2007), Malik et al. (2013) or Dobbs et al. (2014). To try to facilitate the recognition of links between research topic and policy interventions, we produced a more fine-grained description based only on the publications touching upon social environment (Fig. 9). This map does indeed show more clearly some areas of policy concern such as diet, physical activity, children obesity or community education. However, research areas focused on some of the proposed policy interventions are not visible – for example advertising, taxation on obesity-promoting foods, transport and planning, nutritional labelling or subsidies to healthy foods.

Similarly, broad areas such as food production that are mentioned in obesity policy studies are missing (e.g. Malik et al. 2013). Let us notice that these studies take a systemic perspective according to which the obesity epidemic is caused by changes in living environments, diets and lifestyles which stem from wider societal transformations in food production, consumption patterns, social stratification, or urbanisation. Although some of the topics in Fig. 9 (detailed science map) have terms related to these social wider issues (e.g. ‘beverages’ in the diet topic, or ‘disparities’ in the social environment topic), other key issues such as food production do not pop up as separate topics.

In summary, this preliminary analysis suggests that most research related to obesity is focused on biology and medicine and only small part of the obesity research portfolio is related to policy agendas, mainly through public health and social science topics. In other words, our analysis would suggest to develop relatively more research on social environments and determinants rather than on metabolism or treatments.

There are reasons to be cautious regarding the robustness of this finding. First, the database used (WoS) is biased towards the biomedical sciences. Second, scientific titles and abstracts may focus on the technical issues and do not describe the societal implication of the article, because they only have a secondary or tertiary link with policy. Third, one may question the notion of alignment between science-supply and societal demand, since scientific knowledge is not necessary for policy interventions – e.g. governments can decide to tax ‘junk food’ without more socio-economic research on food corporations[6].

In spite of these shortcomings, the insights of the study are in accordance with views expressed in the literature on obesity (Malik et al. (2013) or Dobbs et al. (2014). In particular, the findings that science is more focused on biomedical research than on socio-economic factors of obesity is consistent with the view expressed in the famous Lalonde report (1974). This report, commissioned by the Canadian minister of health and welfare, argued that traditional health care system was too focused on medicine and paid too little attention to prevention and promotion of good health. In agreement with the obesity policy studies mentioned above (e.g. Millstone et al., 2006), the report took a systemic view according to which health is influenced by human biology, health care system, lifestyle and the physical and social environment, suggesting a shift of policy emphasis towards the later two spheres. Here, we also found that most obesity related research is focused on biology and treatment or health care, rather than on the lifestyles and social and economic determinants.

Given that the purpose of our methodology is to facilitate expert deliberation on research priorities, we believe that future research on this agenda should focus, first, on studying the similarities and disagreement between the perspective provided by topic maps and stakeholders, and second, on the potential use of this type of quantitative evidence aimed at pluralization of perspectives in decision-making for priority setting (Rafols et al., 2012).

[6] Yet, if research is irrelevant in policy, one may wonder why soda companies (like the tobacco, pharmaceutical or oil industries did) make such substantial efforts to produce evidence supporting their interests (see e.g. Schillinger et al. (2016) on studies regarding relation between sugar-sweetened beverages and obesity or diabetes and Aaron et al. (2017)) on research sponsorship).

## Acknowledgements

We would like to thank Tommaso Ciarli for suggesting us the use of parliamentary questions as one of the possible representations of social needs. From CWTS (University of Leiden), we thank Ed Noyons for suggesting the delineation method and Ludo Waltman for sharing the article level classification system. We thank to Kevin Boyack for fruitful comments and Christine Cherbut for explaining her views on policy options and scientific priorities for tackling obesity. We also thank the two anonymous reviewers for useful comments on an earlier version of this paper.

## Funding and declaration of interest

This research was part of the OST research programme and did not receive any specific grant from funding agency in the public, commercial or non-profit sectors. Ismael Ràfols was supported by the European Commission Marie Curie Integration Grant.

The authors declare that no competing interests exist.

**Appendix A:** Large disciplinary categories used in this study.

The large disciplines we used are the following groups of OST disciplines

| LARGE DISCIPLINE | OST DISCIPLINE |
|---|---|
| BIOLOGY | AGROFOOD |
| | BIOING |
| | BIOCHEM |
| | NEUROSCI |
| | REPR/DEVBIO |
| | BIOTECH/GENE |
| | MICRO/IMMU |
| MEDICAL RES | NUTR/ENDOCR |
| | DIV MED |
| | SUR/GAS/URO |
| | PHARM/TOXICO |
| | CANCER |
| HEALTH | PUBLIC HEALTH |
| PSYCHO & SOCIAL | HEALTH/SOCIAL |
| | PSYCHO |

OST disciplines are derived from Web of Science Categories as specified in the following link: www.obs-ost.fr/sites/default/files/nomenclatures_disciplinaires_0.pdf

**Appendix B**: Titles of the 16 characteristic papers (with topic weight larger than 0.9) for topic U4

| U4 most characteristic articles | |
|---|---|
| 1 | Minimal in-person support as an adjunct to Internet obesity treatment |
| 2 | A pilot church-based weight loss program for African-American adults using church members as health educators: A comparison of individual and group intervention |
| 3 | A motivation-focused weight loss maintenance program is an effective alternative to a skill-based approach |
| 4 | Developing culturally congruent weight maintenance programs for African American church members |
| 5 | Academic incentives for students can increase participation in and effectiveness of a physical activity program |
| 6 | Nutrition communication for a Latino community: Formative research foundations |
| 7 | School climate and the institutionalization of the CATCH program |
| 8 | Improving Weight Loss Outcomes of Community Interventions by Incorporating Behavioral Strategies |
| 9 | Adapting and implementing a long-term nutrition and physical activity curriculum to a rural, low-income, biethnic community |
| 10 | 'Ready. Set. ACTION!' A theater-based obesity prevention program for children: a feasibility study |
| 11 | Process evaluation of a school-based weight gain prevention program: the Dutch Obesity Intervention in Teenagers (DOiT) |
| 12 | Behavioral and Cognitive Effects of a Worksite-Based Weight Gain Prevention Program: The NHF-NRG In Balance-Project |
| 13 | Diabetes Prevention, Weight Loss, and Social Support Program Participants' Perceived Influence on the Health Behaviors of Their Social Support System |
| 14 | Disseminating Health Promotion Practices in After-School Programs Through YMCA Learning Collaboratives |
| 15 | Nutrition education for student community volunteers: A comparative study of two different communication methods |
| 16 | An exploration of the experiences and perceptions of people who have maintained weight loss |

**Appendix C:** Titles on the 26 characteristic papers (with topic weight larger than 0.9) of topic U1

| U1 most characteristic articles | |
|---|---|
| 1 | The Geography of Recreational Open Space: Influence of Neighborhood Racial Composition and Neighborhood Poverty |
| 2 | Area variation in recreational cycling in Melbourne: a compositional or contextual effect? |
| 3 | Is availability of public open space equitable across areas? |
| 4 | Bicycle Use for Transport in an Australian and a Belgian City: Associations with Built-Environment Attributes |
| 5 | Spatial Disparities in the Distribution of Parks and Green Spaces in the USA |
| 6 | Walking to the bus: perceived versus actual walking distance to bus stops for older adults |
| 7 | Income and racial disparities in access to public parks and private recreation facilities |
| 8 | Differences in self-reported health among Asians, Latinos, and non-Hispanic whites: The role of language and nativity |
| 9 | Travel behavior and objectively measured urban design variables: Associations for adults traveling to work |
| 10 | The importance of place of residence: Examining health in rural and nonrural areas |
| 11 | A spatial analysis of health-related resources in three diverse metropolitan areas |
| 12 | Site and neighborhood environments for walking among older adults |
| 13 | Effects of access to public open spaces on walking: Is proximity enough? |
| 14 | Availability of recreational resources in minority and low socioeconomic status areas |
| 15 | Disparities in Urban Neighborhood Conditions: Evidence from GIS Measures and Field Observation in New York City |
| 16 | Neighbourhood Environmental Correlates of Perceived Park Proximity in Montreal |
| 17 | Food store availability and neighborhood characteristics in the United States |
| 18 | A Neighborhood Wealth Metric for Use in Health Studies |
| 19 | Contextualizing nativity status, Latino social ties, and ethnic enclaves: an examination of the 'immigrant social ties hypothesis' |
| 20 | The relationship between segment-level built environment attributes and pedestrian activity around Bogota's BRT stations |
| 21 | Proximity to trails and retail: Effects on urban cycling and walking |
| 22 | Neighborhood racial composition, neighborhood poverty, and the spatial accessibility of supermarkets in metropolitan Detroit |
| 23 | Regional differences in walking frequency and BMI: What role does the built environment play for Blacks and Whites? |
| 24 | McDonald's restaurants and neighborhood deprivation in Scotland and England |
| 25 | Retail Redlining in New York City: Racialized Access to Day-to-Day Retail Resources |
| 26 | Urban form correlates of pedestrian travel in youth: Differences by gender, race-ethnicity and household attributes |

**Complementary material available on Figshare**

- Science corpora : list of WoS-Clarivate Analytics UT keys for the publications of the 6 science corpora doi:10.6084/m9.figshare.4959791

- Questions related to obesity from Members of the European Parliament to the European Commission (October 2009 - February 2014) doi:10.6084/m9.figshare.4958996

- LDAvis Interactive maps (Five files to download in a directory, *index.html* file to open with Firefox)

  Map of 20 topics for corpus C of obesity publications doi:10.6084/m9.figshare.4959098
  Map of 10 topics for sub-corpus SC of obesity publications doi:10.6084/m9.figshare.4959134
  Map of 30 topics for a corpus of MEP's questions about obesity
  doi:10.6084/m9.figshare.4959752